\documentclass[12pt,tightenlines,floats,aps,amsmath,amssymb,nofootinbib,prd,floatfix]{revtex4}

\usepackage{setspace}
\usepackage{subfigure}
\usepackage{amsmath,amssymb,amsfonts,amsthm,mathrsfs}
\usepackage{graphicx,wrapfig}
\usepackage{enumerate}
\usepackage{color,xcolor}
\usepackage{hyperref}
\usepackage{physics}
\hypersetup{
	bookmarks=true,         
	unicode=false,          
	pdftoolbar=true,        
	pdfmenubar=true,        
	pdffitwindow=false,     
	pdfstartview={FitH},    
	pdftitle={Slow-roll inflation in f(R,T) gravity and a modified Starobinsky-like inflationary model},    
	pdfauthor={Mauricio Gamonal},     
	pdfkeywords={inflation, modified gravity, f(R,T), cosmology}, 
	pdfnewwindow=true,      
	colorlinks=true,       
	linkcolor=red,          
	citecolor=blue,        
	filecolor=cyan,         
	urlcolor=blue        
}

\begin{document}

\title{Slow-roll inflation in $f(R,T)$ gravity and a modified Starobinsky-like inflationary model}

\author{Mauricio Gamonal$^{\dagger}$}
\affiliation{
Instituto de F\'isica, Facultad de F\'isica, Pontificia Universidad Cat\'olica de Chile, Av. Vicu\~na Mackenna 4860, Santiago, Chile.\\
$^\dagger$\url{mfgamonal@uc.cl}
}
\date{\today}
\setlength\parindent{0pt}

\begin{abstract}

\noindent In this work, we studied the slow-roll approximation of cosmic inflation within the context of $f(R,
T)$ gravity, where $R$ is the scalar curvature, and $T$ is the trace of the energy-momentum tensor. By choosing a minimal coupling between matter and gravity, we obtained the modified slow-roll parameters, the scalar spectral index ($n_s$), the tensor spectral index ($n_{\textrm{T}}$), and the tensor-to-scalar ratio ($r$). We computed these quantities for a general power-law potential, Natural \& Quartic Hilltop inflation, and the Starobinsky model, plotting the trajectories on the $(n_s,r)$ plane. We found that one of the parameters of Natural/Hilltop models is non-trivially modified. Besides, if the coupling is in the interval $-0.5<\alpha<5.54$, we concluded that the Starobinsky-like model predictions are in good agreement with the last Planck measurement, but with the advantage of allowing a wide range of admissible values for $r$ and $n_{\textrm{T}}$.

\end{abstract}
\maketitle
\section{Introduction}
\label{sec:Introduction}

Throughout the last decades, several studies have been carried out to describe the cosmological dynamics of the Universe, both on a theoretical and observational level, and all of them seem to be in good agreement with the standard cosmological theory, i.e., the $\Lambda$CDM model within the framework of General Relativity (GR) \cite{Ferreira:2019}. Moreover, the measurement of the Cosmic Microwave Background (CMB), according to the observations of COBE, WMAP, and Planck, has granted us an enormous amount of information about the cosmos. However, the confirmation of a spatially flat and isotropic Universe seems unfavorable (flatness and horizon problems) regarding the standard cosmological model \cite{Coley:2019}. A possible solution to these problems can be found by considering an epoch of exponential expansion in the primordial stage of the Universe, known as \textit{cosmic inflation}, which theoretical framework was developed forty years ago by the contributions of many authors, e.g., Starobinsky \cite{Starobinsky:1980}, Guth \cite{Guth:1981}, Linde \cite{Linde:1982} and Albrecht \& Steinhardt \cite{AlbrechtSteinhardt:1982}. The simplest and the most popular way to study the inflationary scenario is by considering a scalar field, i.e., the inflaton, under the action of a specific potential, and imposing the so-called \textit{slow--roll approximation}, where kinetic terms of the inflaton are usually neglected. Several types of potentials have been used to describe the inflationary dynamics, and their predictions have been tested and severely constrained, like the almost scale-invariant power spectrum for density perturbations, which has been verified by the CMB anisotropy measurements \cite{Martin:2015,Planck:2018}.\\

Nonetheless, although the classical theory of General Relativity is by far the most important and accurate model of gravity that we currently have \cite{Will:2005}, the requirement of a dark sector, e.g., dark energy and dark matter, to fit the cosmological data, has been taken as one of the main motivations to study alternative models of gravity. Among the theories that attempt to solve these puzzles, we will consider the approach of $f(R,T)$ gravity, a model proposed by Harko et al. \cite{Harko:2011} (do not confuse with $f(\mathbb{T})$ modified teleparallel gravity \cite{Ferraro:2006,Golovnev:2020}) and which considers an action described by a function of $R$ (the scalar curvature) and $T$ (the trace of the energy-momentum tensor). Different topics of contemporary cosmology have been widely studied within the framework of the $f(R,T)$ theory, for example: Cosmological models and Solar System consequences \cite{Shabani:2013,Shabani:2014}, Scalar perturbations \cite{Alvarenga:2013}, Thermodynamics \cite{Sharif:2012}, The dark energy \cite{Sun:2015} or dark matter problems \cite{Zaregonbadi:2016}, a general description of the FLRW cosmology \cite{Myrzakulov:2012}, the propagation of Gravitational Waves \cite{Alves:2016,Sharif:2019} or the Hamiltonian formalism in Quantum Cosmology \cite{Xu:2016}. In that sense, a recent work addressed cosmic inflation triggered by a perfect fluid and by a specific quadratic potential \cite{Bhattacharjee:2020_inf}. However, here we aim to develop an exhaustive analysis of the slow-roll approximation to provide a general result, valid for any potential, giving a more extensive view of cosmic inflation in $f(R,T)$ gravity. With such a result, we will be able to determine the modification to the cosmological parameters and identify the physically meaningful corrections due to the action of the $f(R,T)$ theory.\\

The manuscript is structured as follows: In Sec. \ref{sec:Standard cosmology and Inflationary parameters}, we review the slow-roll inflation in General Relativity and introduce the slow-roll parameters, the observable spectral indices, and other cosmological quantities of interest. In Sec. \ref{sec:Review of f(R,T) gravity} we provide the more general equations of $f(R,T)$ gravity. After taking a particular functional form of $f(R, T)$, we will carefully study the slow-roll approximation in the modified model in Sec. \ref{sec:Analysis of two specific examples}. With the slow-roll parameters' corrected expressions, we apply these results to several inflationary models in Sec. \ref{sec:Inflationary_models}. Finally, in Sec. \ref{sec:discussion}, we present a summary and the conclusions of this work.

\section{Slow--Roll Inflation in General Relativity}
\label{sec:Standard cosmology and Inflationary parameters}

The classical theory of General Relativity (GR), developed by Einstein \cite{Einstein:1916} has been the most accurate model to study the gravitational interaction, and it has been experimentally verified several times. The equations of motion for the full theory, in the presence of matter, can be derived from the following action,
\begin{equation}
\label{eq:actionEH}
S_{\textrm{GR}} = \int   \left(\frac{R}{2\kappa} + \mathcal{L}_{m}\right) \sqrt{-g} \dd[4]{x},
\end{equation}
where $R$ is the scalar curvature (the trace of the Ricci tensor $R_{\mu\nu}$, i.e. $R = g^{\mu\nu}R_{\mu\nu}$), $\mathcal{L}_m$ is the matter lagrangian, $g$ is the determinant of the metric tensor, i.e. $g=\det(g_{\mu\nu})$, and $\kappa = 8\pi G$, where $G$ is the Newtonian constant of gravitation and where we used natural units such that $c=\hbar=1$. Throughout this paper, we will use the $(-+++)$ signature for the metric tensor. By applying the action principle, i.e., $\delta S_{\textrm{GR}}=0$, we obtain the Einstein field equations,
\begin{equation}
\label{eq:EFE}
R_{\mu\nu} - \frac{1}{2} R g_{\mu\nu} = \kappa T_{\mu\nu} ,
\end{equation}
where $T_{\mu\nu}$ are the components of the energy--momentum tensor, defined as,
\begin{equation}
\label{eq:Stress-energy tensor}
T_{\mu\nu} \equiv - \frac{2}{\sqrt{-g}} \fdv{(\sqrt{-g}\mathcal{L}_m)}{g^{\mu\nu}} = g_{\mu\nu}\mathcal{L}_m - 2 \fdv{\mathcal{L}_m}{g^{\mu\nu}}.
\end{equation}

A few years after the derivation of the field equations, the modern approach to the study of cosmology started with the construction of a metric that describes an isotropic and homogeneous spacetime, assuring the Cosmological Principle \cite{Friedmann:1924,Lemaitre:1931,Robertson:1935,Walker:1937}. This metric is currently known as the Friedmann--Lemaitre--Robertson--Walker (FLRW) and in spherical coordinates, is given by the following expression,
\begin{equation}
\label{eq:FLRW}
\dd{s}^2 = -\dd{t}^2 + a(t)^2 \qty(\frac{\dd[2]{r}}{1-Kr^2} + r^2 \dd{r}^2 (\dd{\theta}^2+\sin[2](\theta)\dd{\phi}^2)),
\end{equation}
where $a(t)$ is a dimensionless function of time known as the scale factor, and $K$ is the Gaussian curvature of space and the metric. In this work, we will consider a Universe with a flat geometry, i.e., $K=0$, as the cosmological measurements seem to show. When $K=0$, and if we normalize the scale factor such that at the present epoch $t_0$, reads $a(t_0) = 1$, the radial coordinate $r$ and the cosmic time $t$ are comoving coordinates.\\

The simplest inflationary scenario can be induced by the inclusion of a spatially homogeneous scalar field called \textit{inflaton}, denoted by $\varphi =\varphi(t)$, which can be introduced by a Lagrangian of the form,
\begin{equation}
\label{eq:lagrangian_scalar_field}
\mathcal{L}_m^{(\varphi)} = -\frac{1}{2}g^{\mu\nu} \partial_\mu \varphi \partial_\nu \varphi - V(\varphi) =  \frac{1}{2} \dot{\varphi}^2 - V(\varphi),
\end{equation}
where $V(\varphi)$ is some potential. Therefore, the components of the energy-momentum tensor can be computed from \eqref{eq:Stress-energy tensor} and read,
\begin{equation}
T_{\mu\nu}^{(\varphi)} = \partial_\mu \varphi \partial_\nu \varphi + g_{\mu\nu} \qty(\frac{1}{2} \dot{\varphi}^2 - V(\varphi)),
\end{equation}
which can be expressed as a perfect fluid with energy density $\rho_{\varphi}$ and pressure $p_{\varphi}$, 
\begin{equation}
\label{eq:density_and_pressure_GR}
T_{00}^{(\varphi)}  =\frac{\dot{\varphi}^2}{2} + V(\varphi) = \rho_{\varphi},\quad T_{ij}^{(\varphi)} = \qty(\frac{\dot{\varphi}^2}{2} - V(\varphi)) g_{ij}  =p_{\varphi} g_{ij} .
\end{equation}
Moreover, the trace of the energy-momentum tensor is given by,
\begin{equation}
\label{eq:trace_T_GR}
T^{(\varphi)} = g^{\mu\nu}T_{\mu\nu}^{(\varphi)} = \dot{\varphi}^2 - 4V(\varphi).
\end{equation}
Thus, if we introduce equation \eqref{eq:density_and_pressure_GR} into the 00 component of the equation \eqref{eq:EFE}, we get 
\begin{equation}
\label{eq:1stFriedmannEq_GR}
H^2 = \frac{\kappa \rho_\varphi}{3} = \frac{\kappa}{3} \qty(\frac{\dot{\varphi}^2}{2} + V(\varphi)), 
\end{equation}
commonly known as the first Friedmann equation and where we have defined the Hubble parameter as $H \equiv \dot{a}/a$. On the other hand, the trace of the field equations \eqref{eq:EFE} reads $R = - \kappa T$. Hence, by rearranging the trace equation, we obtain the second Friedmann equation (or acceleration equation),
\begin{equation}
\label{eq:2ndFriedmannEq_GR}
\frac{\ddot{a}}{a} = - \frac{\kappa}{6} (3p_\varphi+\rho_\varphi) = - \frac{\kappa}{3} \qty( \dot{\varphi}^2 - V(\varphi)  ).
\end{equation}
Furthermore, from the definition of the Hubble parameter, the continuity equation for the energy density and the pressure reads,
\begin{equation}
\label{eq:continuity_equation_GR}
\dot{\rho}_{\varphi} + 3H\qty(\rho_\varphi + p_\varphi) = 0,
\end{equation}
which is, as a matter of fact, the $\mu=0$ component of the conservation of the energy-momentum tensor, i.e., $\nabla_\nu T^{\mu\nu} = 0$. Moreover, by inserting \eqref{eq:density_and_pressure_GR} into \eqref{eq:continuity_equation_GR}, we get the Klein--Gordon equation for the inflaton field (which can also be obtained from a variation on the action with respect to $\varphi$), given by the expression,
\begin{equation}
\label{eq:Klein-Gordon_GR}
\ddot{\varphi} +3H\dot{\varphi} + V_{,\varphi} = 0, 
\end{equation}
where $V_{,\varphi} = \dv{V}{\varphi}$. Usually, the inflationary scenario at the early stages of the Universe is characterized by a quasi--exponential rate of expansion, i.e., $\dv{(H^{-1})}{t}\ll 1$, which implies the slow--roll condition,
\begin{equation}
\label{eq:slow--roll_condition}
\dot{\varphi}^2 \ll V(\varphi).
\end{equation}
Therefore, we can define the first slow--roll parameter, denoted by $\epsilon$, as
\begin{equation}
\label{eq:slow-roll_parameters_GR}
\epsilon = -\frac{\dot{H}}{H^2}=\frac{3\dot{\varphi}^2}{\dot{\varphi}^2+2V(\varphi)},
\end{equation} 
such that the minimum requirement to develop inflation is $\abs{\epsilon} \ll 1$. If we apply the slow-roll approximation \eqref{eq:slow--roll_condition} and use the Friedmann equations, we can define at first order a similar slow-roll parameter, denoted by $\epsilon_{\textrm{V}}$ which depends only in the potential $V(\varphi)$,
\begin{equation}
\label{eq:epsilon_V_GR}
\epsilon \approx \frac{3\dot{\varphi}^2}{2V(\varphi)} = \frac{1}{2\kappa} \qty(\frac{V_{,\varphi}}{V})^2 \equiv \epsilon_{\textrm{V}}, 
\end{equation}
If we take the derivative with respect to cosmic time, we can define the second slow-roll parameter, denoted by $\eta$, which guarantees the slow variation of $\epsilon$ in time, 
\begin{equation}
\dot{\epsilon} = 2\frac{\dot{H}^2}{H^3} - \frac{\ddot{H}}{H^2} = 2H\epsilon (\epsilon - \eta), \quad \eta \equiv - \frac{\ddot{\varphi}}{H\dot{\varphi}}.
\end{equation}
Similarly to the case of $\epsilon_{\textrm{V}}$, we can define an $\eta_{\textrm{V}}$ that depends only on the potential. Using \eqref{eq:Klein-Gordon_GR} and the Friedmann equations we have,
\begin{equation}
\label{eq:eta_V_GR}
\eta_{\textrm{V}} = \eta + \epsilon \approx \frac{1}{\kappa} \qty(\frac{V_{,\varphi\varphi}}{V}) ,
\end{equation}
where $V_{,\varphi\varphi}=\dv[2]{V}{\varphi}$. These slow--roll parameters approximately describe the dynamics of inflation and the observational features of different models. In fact, we can write the following spectral indices in terms of the slow--roll parameters \cite{Liddle:2000,Piattela:2018},
\begin{subequations}
\label{eq:spectral_indices_GR}
\begin{align}
n_\textrm{s}-1 &= \dv{\ln(\Delta_{\textrm{S}}^2)}{\ln(k)} = -4\epsilon + 2 \eta \approx - 6\epsilon_{\textrm{V}} + 2\eta_{\textrm{V}} \\
n_\textrm{T} &= \dv{\ln(\Delta_{\textrm{T}}^2)}{\ln(k)} = -2 \epsilon \approx -2\epsilon_{\textrm{V}} \\
r_{*} &= \frac{\Delta_{\textrm{T}}^2(k_{*})}{\Delta_{\textrm{S}}^2(k_{*})} = 16\epsilon \approx 16 \epsilon_{\textrm{V}}, 
\end{align}
\end{subequations}
where $\Delta_{\textrm{S}}$ and $\Delta_{\textrm{T}}$ are the dimensionless scalar and tensor power spectrum, respectively; $n_s$ is the scalar spectral index, $n_{\textrm{T}}$ is the tensor spectral index and $r_{*}$ is the tensor--to--scalar ratio at the scale $k_{*}$. Here we have neglected the running of the spectral indices. Another important quantity is the number of e--folds, defined as $N=\ln(a)$, which measures the amount of spacetime expansion. The slow--roll approximation yields a $N$ given by,
\begin{eqnarray}
\label{eq:e-folds_GR}
N = \int_{t_1}^{t_2} H\dd{t} = \int_{\varphi_{\textrm{end}}}^{\varphi} \frac{H}{\dot{\varphi}} \dd{\varphi} \approx \kappa \int_{\varphi_{\textrm{end}}}^{\varphi} \frac{V(\varphi')}{V_{\varphi}(\varphi')} \dd{\varphi'},
\end{eqnarray}
where $\varphi_{\textrm{end}}$ is the inflaton value at the end of inflation, i.e. when $\epsilon_{\textrm{V}}$ or $\eta_{\textrm{V}}$ is close to 1, and the integral upper limit usually refers to the value of $\varphi$ at the horizon crossing. In summary, knowing the functional form of the potential $V(\varphi)$ would yield predictions susceptible to experimental verification by measuring the primordial power spectrum. 

\section{$f(R,T)$ gravity}
\label{sec:Review of f(R,T) gravity}

Here, we will discuss the modified model of gravity introduced by Harko et al. \cite{Harko:2011}, known as $f(R,T)$ gravity, is described by the following action,
\begin{equation}
\label{eq:f(R,T)_Action}
S = \int \qty(\frac{f(R,T)}{2\kappa} + \mathcal{L}_m) \sqrt{-g} \dd[4]{x},
\end{equation}
where $f(R,T)$ is an arbitrary function of the scalar curvature, $R$, and the trace of the energy-momentum tensor, $T$, and where $\mathcal{L}_m$ is the matter lagrangian, such that $T_{\mu\nu}$ is defined as in the equation \eqref{eq:Stress-energy tensor}. By varying the action \eqref{eq:f(R,T)_Action} with respect to the metric, we obtain the $f(R,T)$ gravity field equations,
\begin{equation}
\label{eq:f(R,T)_EoM}
\pdv{f}{R} R_{\mu\nu} - \frac{1}{2} f(R,T) g_{\mu\nu} + (g_{\mu\nu}\square - \nabla_\mu \nabla_\nu) \pdv{f}{R} = \kappa T_{\mu\nu} - \pdv{f}{T} \qty(T_{\mu\nu} + \Theta_{\mu\nu}) ,
\end{equation} 
where $\square \equiv \nabla^\mu \nabla_\mu$, while $\nabla_\mu$ is the covariant derivative, and  $\Theta_{\mu\nu}$ is defined as
\begin{equation}
\label{eq:f(R,T)_Theta}
\Theta_{\mu\nu} \equiv g^{\alpha\beta} \fdv{T_{\alpha\beta}}{g^{\mu\nu}} = - 2 T_{\mu\nu}  +g_{\mu\nu} \mathcal{L}_m - 2 g^{\alpha\beta} \frac{\delta^2\mathcal{L}_m}{\delta g^{\mu\nu}\delta g^{\alpha\beta}}.
\end{equation}
Note that when $f(R,T)$ does not depend on $T$, the equations of motion  \eqref{eq:f(R,T)_EoM} reduce to the well--known $f(R)$ gravity. In this work we will assume that $f(R,T) = R + 2\kappa\alpha  T$, where $\alpha$ is a real constant. This is the most studied model of $f(R,T)$ gravity and its viability has been investigated in numerous works \cite{Momeni:2011,Moraes:2015_SE,Yousaf:2016,Moraes:2016,Moraes:2017_WHs,Shabani:2017,Santos:2018nqb,Yousaf:2019,Taser:2020,Maurya:2020ebd,Yadav:2020,Bhattacharjee:2020a,Bhattacharjee:2020b,Tiwari:2021}. In this particular model, the action of gravity reads,
\begin{equation}
\label{eq:f(R,T)_Action_Final}
S = \int \qty(\frac{R}{2\kappa} + \alpha T ) \sqrt{-g}\dd[4]{x} + \int \mathcal{L}_m \sqrt{-g} \dd[4]{x},
\end{equation}
and from \eqref{eq:f(R,T)_EoM}, the equations of motion can take the following form,
\begin{equation}
\label{eq:f(R,T)_EoM_effective_T}
R_{\mu\nu} - \frac{1}{2} g_{\mu\nu} R = \kappa T_{\mu\nu}^{(\textrm{eff})},
\end{equation}
where we have defined an effective energy-momentum tensor as,
\begin{equation}
\label{eq:Effective_T_munu}
T_{\mu\nu}^{(\textrm{eff})} \equiv T_{\mu\nu} - 2\alpha\qty(T_{\mu\nu}- \frac{1}{2}Tg_{\mu\nu}+\Theta_{\mu\nu} ).
\end{equation}

It is clear from the last expression that when $\alpha \to 0$, we completely recover General Relativity dynamics.  

\section{Slow--Roll inflation in $f(R,T)$ gravity}
\label{sec:Analysis of two specific examples}
Now we will focus our study on the case in which $\mathcal{L}_m$ is given by \eqref{eq:lagrangian_scalar_field}, i.e. a scalar field minimally coupled to $f(R,T)$ gravity. Similar approaches were studied in \cite{Bhattacharjee:2020_inf} and \cite{Aygun:2018_sca}. However, unlike these works, we want to analyze the slow-roll approximation exhaustively and use it in different inflationary models. For a single and spatially homogeneous scalar field we have,
\begin{equation}
\Theta_{\mu\nu}^{(\varphi)} = -2T_{\mu\nu}^{(\varphi)} + g_{\mu\nu} \mathcal{L}_m^{(\varphi)} =-2\partial_\mu \varphi \partial_\nu \varphi - g_{\mu\nu} \qty(\frac{1}{2} \dot{\varphi}^2 - V(\varphi)  ),
\end{equation}
with,
\begin{equation}
\Theta_{00}^{(\varphi)} = -T^{(\varphi)}_{00}-\dot{\varphi}^2,\quad \Theta_{ij}^{(\varphi)} = - T_{ij}^{(\varphi)},
\end{equation}
where $T_{\mu\nu}^{(\varphi)}$ is given by \eqref{eq:density_and_pressure_GR}. Thus, the trace of $\Theta_{\mu\nu}^{(\varphi)}$ can be easily computed,
\begin{equation}
\Theta^{(\varphi)} = g^{\mu\nu} \Theta_{\mu\nu}^{(\varphi)} = 4V(\varphi).
\end{equation}
Now we can compute the components of the effective energy-momentum tensor defined in \eqref{eq:Effective_T_munu}. For a homogeneous inflaton field, the energy-momentum tensor takes a diagonal form, from where we can define an effective energy density and pressure,
\begin{align}
\label{eq:Effective_rho}
T_{00}^{(\textrm{eff})} &= \frac{1}{2}\dot{\varphi}^2 (1+2\alpha) + V(\varphi) (1+4\alpha) \equiv \rho_{\varphi}^{(\textrm{eff})} \\
T_{ij}^{(\textrm{eff})} &= \qty(\frac{1}{2}\dot{\varphi}^2 (1+2\alpha) - V(\varphi) (1+4\alpha) ) g_{ij} \equiv p_{\varphi}^{(\textrm{eff})} g_{ij}, \label{eq:Effective_pressure}
\end{align}
and $T^{(\textrm{eff})}_{\mu\nu} = 0$ for $\mu\neq \nu$. From these expressions, we can define an effective equation of state as the ratio of the pressure and the energy density,

\begin{equation}
\label{eq:f(R,T)_EoS}
w^{(\textrm{eff})} \equiv \frac{p_{\varphi}^{(\textrm{eff})}}{\rho_{\varphi}^{\textrm{(eff)}}} = \frac{\dot{\varphi}^2(1+2\alpha)-2V(\varphi)(1+4\alpha)}{\dot{\varphi}^2(1+2\alpha)+ 2V(\varphi)(1+4\alpha)}.
\end{equation}

Additionally, the trace of $T^{(\textrm{eff})}_{\mu\nu}$ can also be computed,
\begin{equation}
\label{eq:Effective_T_munu_Trace}
T^{(\textrm{eff})} = g^{\mu\nu}T^{(\textrm{eff})}_{\mu\nu} = \dot{\varphi}^2 (1+2\alpha) - 4V(\varphi) (1+4\alpha) = 3p_{\varphi}^{(\textrm{eff})} - \rho_{\varphi}^{(\textrm{eff})} = (3w^{(\textrm{eff})}-1)\rho_{\varphi}^{(\textrm{eff})}.
\end{equation}
By using the above expressions, in addition to \eqref{eq:f(R,T)_EoM_effective_T}, we can obtain the generalized Friedmann equations for our particular model of $f(R,T)$ gravity. From them, we can also get an expression for $\dot{H}$. The resulting equations read,
\begin{align}
H^2 &= \frac{\kappa}{3} \rho_{\varphi}^{(\textrm{eff})} = \frac{\kappa}{3} \qty(\frac{\dot{\varphi}^2}{2} (1+2\alpha) + V(\varphi) (1+4\alpha) ), \label{eq:f(R,T)_Friedmann_1}\\
\frac{\ddot{a}}{a} &= - \frac{\kappa}{6} (3p_{\varphi}^{(\textrm{eff})} + \rho_{\varphi}^{(\textrm{eff})} ) =  - \frac{\kappa}{3} \qty( \dot{\varphi}^2(1+2\alpha) - V(\varphi)(1+4\alpha  )) \label{eq:f(R,T)_Friedmann_2}\\
\dot{H} &= \frac{\ddot{a}}{a} - H^2 = -\frac{\kappa}{2} (p_{\varphi}^{(\textrm{eff})}  + \rho_{\varphi}^{(\textrm{eff})} ) = - \frac{\kappa \dot{\phi}^2}{2} (1+2\alpha) .\label{eq:f(R,T)_H_dot}
\end{align}
Furthermore, we can find a continuity equation for $\rho_{\varphi}^{(\textrm{eff})}$ and $p_{\varphi}^{(\textrm{eff})}$ through a similar procedure that gave us equation \eqref{eq:continuity_equation_GR}. By taking a time derivative on \eqref{eq:f(R,T)_Friedmann_1} and replacing \eqref{eq:f(R,T)_H_dot}, we obtain the modified Klein--Gordon equation,
\begin{equation}
\label{eq:f(R,T)_Klein-Gordon}
\ddot{\varphi}(1+2\alpha)+3H\dot{\varphi} (1+2\alpha) + V_{,\varphi}(1+4\alpha) = 0.
\end{equation}
From the previous analysis we observe that the structure of the cosmological equations in this particular model of $f(R,T)$ gravity is conserved as we replace $p_{\varphi} \to p_{\varphi}^{(\textrm{eff})}$ and $\rho_{\varphi}\to \rho_{\varphi}^{(\textrm{eff})}$. Therefore, the corrections to the inflationary dynamics are be enclosed within the corrections to energy density and pressure due to $f(R,T)$ gravity. These corrections seem to show that the conserved energy--momentum tensor is actually $T_{\mu\nu}^{(\textrm{eff})}$, as it can be noted from \eqref{eq:f(R,T)_EoM_effective_T}. Taking into account this idea, we will proceed to specify the slow--roll condition according to this model and, then, to compute the corrections to the slow-roll parameters $\tilde{\epsilon}_{\textrm{V}}$ and $\tilde{\eta}_{\textrm{V}}$, where we will denote with a tilde the parameters within the framework of $f(R,T)$ gravity. The first slow--roll parameter is given by,
\begin{equation}
\label{eq:f(R,T)_epsilon}
\tilde{\epsilon} = -\frac{\dot{H}}{H^2} = \frac{3}{2} \frac{\dot{\varphi}^2(1+2\alpha)}{\qty(\frac{\dot{\varphi}^2}{2}(1+2\alpha)+V(\varphi)(1+4\alpha))}.
\end{equation}
An inflationary evolution of the Universe requires that $\abs{\tilde{\epsilon}}\ll1$. Hence, the slow-roll condition for this model becomes,
\begin{equation}
\label{eq:f(R,T)_Slow-roll_condition_1}
\dot{\varphi}^2 (1+2\alpha) \ll V(\varphi) (1+4\alpha).
\end{equation}
On the other hand, we can show that the second slow-roll parameter is not corrected,
\begin{equation}
\dot{\tilde{\epsilon}} = 2H\tilde{\epsilon}^2 + \frac{\kappa \dot{\varphi} \ddot{\varphi}(1+2\alpha)}{H^2} = 2H\tilde{\epsilon}^2 + 2\tilde{\epsilon} \frac{\ddot{\varphi}}{\dot{\varphi}} = 2H\tilde{\epsilon}(\tilde{\epsilon}-\tilde{\eta}),\quad \tilde{\eta} = -\frac{1}{H}\frac{\ddot{\varphi}}{\dot{\varphi}} .
\end{equation}
Thus, the conditions $\abs{\tilde{\epsilon}}\ll 1$ and $\abs{\tilde{\eta}} \ll 1$ imply that we can neglect the second order derivative in \eqref{eq:f(R,T)_Klein-Gordon}, such that the Klein-Gordon equation becomes,
\begin{equation}
\label{eq:f(R,T)_slow-roll_condition_2}
3H\dot{\varphi} (1+2\alpha) \approx - V_{,\varphi} (1+4\alpha) .
\end{equation}
By using this expression, and by applying the slow-roll condition \eqref{eq:f(R,T)_Slow-roll_condition_1} into \eqref{eq:f(R,T)_Friedmann_1}, we can find the correction to the first potential slow-roll parameter due to $f(R,T)$ gravity, which we denote as $\tilde{\epsilon}_{\textrm{V}}$
\begin{equation}
\label{eq:f(R,T)_epsilon_V}
\tilde{\epsilon} \approx \frac{3\dot{\varphi}^2(1+2\alpha)}{2V(\varphi)(1+4\alpha)} = \frac{3}{2} \frac{1+2\alpha}{V(1+4\alpha)} \qty(\frac{V_{,\varphi}}{3H}\qty(\frac{1+4\alpha}{1+2\alpha}))^2 = \frac{1}{2\kappa} \qty(\frac{V_{,\varphi}}{V})^2 \qty(\frac{1}{1+2\alpha}) \equiv \tilde{\epsilon}_{\textrm{V}}.
\end{equation}
Additionally, if we take a time derivative on \eqref{eq:f(R,T)_slow-roll_condition_2}, the term $\ddot{\varphi}$ can be expressed in terms of $V_{,\varphi\varphi}$, by using the chain rule such that $\dot{V}=V_{\varphi}\dot{\varphi}$. Therefore, the second slow--roll parameter is given by,
\begin{equation}
\tilde{\eta} = -\frac{1}{H}\frac{\ddot{\varphi}}{\dot{\varphi}} = \frac{1}{3H^2} \frac{V_{,\varphi\varphi}(1+4\alpha)}{1+2\alpha} + \frac{\dot{H}}{H^2},
\end{equation}
and by using \eqref{eq:f(R,T)_Friedmann_1}, we can obtain the corrected expression for $\tilde{\eta}_V$ in the slow-roll approximation, which reads
\begin{equation}
\label{eq:f(R,T)_eta_V}
\tilde{\eta}_{\textrm{V}} \equiv \tilde{\epsilon} + \tilde{\eta} \approx \frac{1}{\kappa}\qty(\frac{V_{,\varphi\varphi}}{V}) \qty(\frac{1}{1+2\alpha}).
\end{equation}
The number of e-folds are also modified in this model. The value can be computed from its definition in equation \eqref{eq:e-folds_GR}, also with \eqref{eq:f(R,T)_slow-roll_condition_2} and \eqref{eq:f(R,T)_Friedmann_1}, such that we get,
\begin{eqnarray}
\label{eq:f(R,T)_e-folds_definition}
\tilde{N} = \int \frac{H}{\dot{\varphi}} \dd{\varphi} \approx \kappa(1+2\alpha) \int_{\varphi_{\textrm{end}}}^{\varphi} \frac{V}{V_{,\varphi}} \dd{\varphi}.
\end{eqnarray}
It is noteworthy that these expressions are independent of the form of the potential $V(\varphi)$, and hence, the results presented can be interpreted as a generalization of the slow-roll approximation for the model $f(R,T) = R+2\kappa \alpha T$ of gravity. It should also be noted that certain values for $\alpha$ are problematic since choosing them would cause the slow-roll parameters to blow up or become zero. In general, we will consider that $\alpha > -1/2$ as a minimum requirement to have well-defined parameters.\\

Another essential feature that we can infer from these results is that this model is entirely equivalent to a particular case of a scalar-tensor model of gravity. For instance, the most general action of scalar-tensor gravity is given by \cite{Faraoni:2004pi},
\begin{equation}
\label{eq:scalar-tensor}
S = \int \qty(f(\phi)R- \frac{\omega(\phi)}{\phi}  \partial^{\alpha}\phi \partial_\alpha \phi - v(\phi) )\sqrt{-g} \dd[4]{x}.
\end{equation}
Hence, it can be seen that the action \eqref{eq:f(R,T)_Action} is equivalent to the above action \eqref{eq:scalar-tensor} if $f(\phi) = 1$, $\omega(\phi) = (1+2\alpha)(\phi/2)$ and $v(\phi)=(1+4\alpha)V(\phi)$, where $V(\phi)$ is the original potential. In fact, a further equivalence can be deduced from the above analysis. The action \eqref{eq:f(R,T)_Action_Final} can be recast as a scalar field minimally coupled to General Relativity, i.e.~the Einstein frame~\cite{Postma:2014},
$$ S = \int  \qty(\frac{R}{2\kappa} - \frac{1}{2}\partial^\alpha\bar{\varphi} \partial_{\alpha} \bar{\varphi} - \bar{V}_{\textrm{eff}}(\bar{\varphi}) )\sqrt{-g}\dd[4]{x}, $$ 
if we perform the following transformations over the field and the potential,
\begin{equation}
\label{eq:f(R,T)_effective_potential}
\bar{\varphi} \to \sqrt{1+2\alpha}\;\varphi ,\quad \bar{V}_{\textrm{eff}}(\bar{\varphi}) \to (1+4\alpha) V\qty(\frac{\bar{\varphi}}{\sqrt{1+2\alpha}}),
\end{equation}
where $V$ is the potential initially considered in the original lagrangian \eqref{eq:lagrangian_scalar_field}.

\section{Inflationary models in $f(R,T)$ gravity and a comparison with Planck 2018 constraints}
\label{sec:Inflationary_models}

\subsection{Power Law Potentials}

As the first example of an inflationary scenario, we will take one of the simplest models by taking a power-law potential of the form,
\begin{equation}
\label{eq:power_law_potential}
V(\varphi) = \lambda\varphi^n,
\end{equation}
where $\lambda$ is a coupling constant. From the definitions of $\epsilon_{\textrm{V}}$ and $\eta_{\textrm{V}}$, in equations \eqref{eq:epsilon_V_GR} and \eqref{eq:eta_V_GR} respectively, we obtain the following slow--roll parameters,
\begin{equation}
\epsilon_{\textrm{V}} = \frac{n^2}{2\kappa \varphi^2}, \quad \eta_{\textrm{V}} = \frac{n(n-1)}{\kappa \varphi^2}.
\end{equation}
As inflation ends when $\epsilon_{\textrm{V}}(\varphi_{\textrm{end}}) = 1$, we have, $\varphi_{\textrm{end}} = \frac{n}{\sqrt{2\kappa}}$. Therefore, the number of e--folds can be computed from \eqref{eq:e-folds_GR}, 
\begin{equation}
N = \frac{\kappa}{2n} \qty(\varphi^2 - \frac{n^2}{2\kappa})
\end{equation}
Thus, the slow-roll parameters can be expressed in terms of the number of e-folds as follows,
\begin{equation}
\epsilon_{\textrm{V}} = \frac{n}{4N+n} , \quad \eta_{\textrm{V}} = \frac{2(n-1)}{4N+n}.
\end{equation}
The spectral indices for this model can be found from \eqref{eq:spectral_indices_GR},
\begin{subequations}
\label{eq:spectral_indexes_power_law_GR}
\begin{align}
n_{\textrm{S}} - 1 &\approx 2\eta_{\textrm{V}} - 6\epsilon_{\textrm{V}} = - \frac{2(n+2)}{4N+n}\\
n_{\textrm{T}} &\approx -2\epsilon_{\textrm{V}} = -\frac{2n}{4N+n}\\
r &\approx 16 \epsilon_{\textrm{V}} = \frac{16n}{4N+n}
\end{align}
\end{subequations}

Now we turn out to compute the corrections to the slow-roll parameters. From \eqref{eq:f(R,T)_epsilon_V} and \eqref{eq:f(R,T)_eta_V}, we obtain,
\begin{align}
\tilde{\epsilon}_{\textrm{V}} &= \qty(\frac{1}{1+2\alpha})\frac{n^2}{2\kappa \varphi^2} \\
\tilde{\eta}_{\textrm{V}} &= \qty(\frac{1}{1+2\alpha}) \frac{n(n-1)}{\kappa \varphi^2}.
\end{align}
Inflation ends when the slow-roll condition is no longer valid, i.e. $\bar{\epsilon}_{\textrm{V}} \approx 1$. Thus, we have
\begin{equation}
\varphi_{\textrm{end}}^2 = \frac{n^2}{2\kappa (1+2\alpha)}.
\end{equation}
Therefore, from \eqref{eq:f(R,T)_e-folds_definition}, the number of e--folds, $\tilde{N}$, is given by the expression,
\begin{equation}
\tilde{N} = (1+2\alpha) \kappa \int_{\varphi_{\textrm{end}}}^{\varphi} \frac{V(\varphi)}{V_{,\varphi}(\varphi)}\dd{\varphi} =  \frac{(1+2\alpha)\kappa}{2n} \qty(\varphi^2-\frac{n^2}{2\kappa(1+2\alpha)}).
\end{equation}
With this, we can express the slow-roll parameters $\tilde{\epsilon}_{\textrm{V}}$ and $\tilde{\eta}_{\textrm{V}}$ as,
\begin{subequations}
	\begin{align}
	\tilde{\epsilon}_{\textrm{V}} &= \frac{n}{4\tilde{N}+n}\\
	\tilde{\eta}_{\textrm{V}} & = \frac{2(n-1)}{4\tilde{N} + n  } 
	\end{align}
\end{subequations}
Therefore, the spectral indices read from \eqref{eq:spectral_indices_GR},
\begin{subequations}
\label{eq:spectral_indices_f(R,T)_power_law}
\begin{align}
\label{eq:spectral_indexes_power_law_f(R,T)}
n_{\textrm{S}} - 1 &= 2\tilde{\eta}_{\textrm{V}} - 6\tilde{\epsilon}_{\textrm{V}} =  - \frac{2(n+2)}{4\tilde{N} +n }\\
n_{\textrm{T}} &= - 2\tilde{\epsilon}_{\textrm{V}} =  -\frac{2n}{4\tilde{N} + n}\\
r&= 16 \tilde{\epsilon}_{\textrm{V}} = \frac{16n}{4\tilde{N} + n} = \frac{8n(1-n_{\textrm{S}})}{n+2} \label{eq:r_power_law_f(R,T)}
\end{align}
\end{subequations}

It is clear that no corrections from $f(R,T)$ gravity are induced to monomial power-law potentials, as the structure of the slow-roll parameters and the spectral indices remain unaltered. The above results seem to differ from the conclusions of \cite{Bhattacharjee:2020_inf}, which indicates that $f(R,T)$ gravity modifies the slow-roll parameters for a quadratic potential in a non-trivial way. We can explain the lack of dependence on $\alpha$ in equations \eqref{eq:spectral_indices_f(R,T)_power_law} from the idea of the effective potential of equation \eqref{eq:f(R,T)_effective_potential}. For a monomial term, for instance, the potential in \eqref{eq:power_law_potential}, we have,
\begin{equation}
\label{eq:effective_potential_power_law}
\bar{V}_{\textrm{eff}} (\bar{\varphi}) = (1+4\alpha)\lambda \qty(\frac{\bar{\varphi}}{\sqrt{1+2\alpha}})^n = \bar{\lambda} \bar{\varphi}^n.
\end{equation}

Thus, the coefficients related to $\alpha$ can be incorporated into a single coupling constant, which determines the strength of the potential, but that does not affect the values of the slow-roll parameters or the spectral indices since they are fixed by the measurements of the CMB amplitudes. This fact implies that $\alpha$ almost have no incidence on the dynamics of an inflationary model triggered by a simple power-law potential and cannot improve the tension between the predictions of power-law potentials and the constraints of the spectral indices measured by Planck 2018 for $50<\tilde{N}<60$ \cite{Planck:2018}. The last assertion does not mean that a different functional form of $f(R,T)$ gravity cannot generate a distinguishable effect for power-law potentials; however, the analysis of these different approaches goes beyond the scope of this work. \\

\subsection{Natural \& Hilltop Inflation}

Now let us apply our results to other types of potentials. To start, we consider Natural inflation \cite{Freese:1990,Adams:1992}, a model whose inflaton field is a pseudo-Nambu-Goldstone boson produced by a spontaneous symmetry breaking in addition to an explicit symmetry breaking, such that the dynamics of a single field is governed by a potential of the form,
\begin{equation}
\label{eq:Natural_Inflation_Pot}
V(\varphi) = \Lambda^4 \qty[1+\cos(\frac{\varphi}{f})],
\end{equation}
where $f$ and $\Lambda$ are mass scales. It has been shown that this potential can drive inflation if $\Lambda \sim M_{\textrm{GUT}} \sim 10^{16}\,\textrm{GeV}$ and $f\sim M_{\textrm{Pl}}=\kappa^{-1/2}$, where $M_{\textrm{Pl}}$ is the reduced Planck mass. Following a similar procedure as before, we can express the slow-roll parameters in terms of the number of e-folds $N$, such that,
\begin{align}
\epsilon_{\textrm{V}} &= \frac{M_{\textrm{Pl}}^2}{2f^2} \frac{\sin[2](\varphi/f)}{\qty[1+\cos(\varphi/f)]^2} =\frac{M_{\textrm{Pl}}^2}{2f^2} \frac{1}{e^{NM_{\textrm{Pl}}^2/f^2}-1}\\
\eta_{\textrm{V}} &= - \frac{M_{\textrm{Pl}}^2}{f^2} \frac{\cos(\varphi/f)}{1+\cos(\varphi/f)} = - \frac{M_{\textrm{Pl}}^2}{2f^2} \frac{e^{NM_{\textrm{Pl}}^2/f^2}-2}{e^{NM_{\textrm{Pl}}^2/f^2}-1}
\end{align}
Thus, the scalar spectral index and the tensor-to-scalar ratio, in the slow-roll approximation, read from \eqref{eq:spectral_indices_GR},
\begin{subequations}
\begin{align}
n_{\textrm{S}} -1 &\approx - \frac{M_{\textrm{Pl}}^2}{f^2}\qty(\frac{e^{NM_{\textrm{Pl}}^2/f^2}+1}{e^{NM_{\textrm{Pl}}^2/f^2}-1})\\
n_{\textrm{T}} &\approx -\frac{M_{\textrm{Pl}}^2}{f^2}\qty(\frac{1}{e^{NM_{\textrm{Pl}}^2/f^2}-1})\\
r &\approx \frac{8M_{\textrm{Pl}}^2}{f^2}\qty(\frac{1}{e^{NM_{\textrm{Pl}}^2/f^2}-1}).
\end{align}
\end{subequations}
Moreover, we can rewrite $r$ in terms of $n_s$, in order to have the representation in the $(r,n_s)$ plane,
\begin{equation}
r(n_s) \approx 4\qty(1-n_s-\frac{M_{\textrm{Pl}}^2}{f^2}).
\end{equation}
Let us analyze if any change is induced by $f(R,T)$ gravity to this model. The corrected slow--roll parameters can be computed following the same recipe as before, which gives
\begin{align}
\tilde{\epsilon}_{\textrm{V}} &= \frac{M_{\textrm{Pl}}^2}{2(1+2\alpha)f^2} \frac{\sin[2](\varphi/f)}{\qty[1+\cos(\varphi/f)]^2} =\frac{M_{\textrm{Pl}}^2}{2(1+2\alpha)f^2} \frac{1}{e^{\frac{\tilde{N}M_{\textrm{Pl}}^2}{(1+2\alpha)f^2}}-1}\\
\tilde{\eta}_{\textrm{V}} &= - \frac{M_{\textrm{Pl}}^2}{(1+2\alpha)f^2} \frac{\cos(\varphi/f)}{1+\cos(\varphi/f)} = - \frac{M_{\textrm{Pl}}^2}{2(1+2\alpha)f^2} \frac{e^{\frac{\tilde{N}M_{\textrm{Pl}}^2}{(1+2\alpha)f^2}}-2}{e^{\frac{\tilde{N}M_{\textrm{Pl}}^2}{(1+2\alpha)f^2}}-1}
\end{align}  
With this, the spectral indices read,
\begin{subequations}
\begin{align}
n_{\textrm{S}} -1 &\approx - \frac{M_{\textrm{Pl}}^2}{(1+2\alpha)f^2}\qty(\frac{e^{\frac{\tilde{N}M_{\textrm{Pl}}^2}{(1+2\alpha)f^2}}+1}{e^{\frac{\tilde{N}M_{\textrm{Pl}}^2}{(1+2\alpha)f^2}}-1})\\
n_{\textrm{T}} &\approx -\frac{M_{\textrm{Pl}}^2}{(1+2\alpha)f^2}\qty(\frac{1}{e^{\frac{\tilde{N}M_{\textrm{Pl}}^2}{(1+2\alpha)f^2}}-1})\\
r &\approx \frac{8M_{\textrm{Pl}}^2}{(1+2\alpha)f^2}\qty(\frac{1}{e^{\frac{\tilde{N}M_{\textrm{Pl}}^2}{(1+2\alpha)f^2}}-1}).
\end{align}
\end{subequations}
From the above expressions we can solve $\tilde{N}$ in terms of $n_{\textrm{S}}$ and obtain the representation in the $(r,n_{\textrm{S}})$ plane, which reads,
\begin{equation}
\label{eq:plane-natural}
r(n_{\textrm{S}}) = 4\qty(1-n_{\textrm{S}} - \frac{M_{\textrm{Pl}}^2}{(1+2\alpha)f^2}).
\end{equation}
It is evident that in Natural inflation there is a non-trivial modification to the values of the inflationary quantities due to the action of $f(R,T)$ gravity. We can summarize all the $\alpha$ contribution to the spectral indices and slow--roll parameters as a correction to the mass scale $f$, i.e., $f\to \sqrt{1+2\alpha}f$, which could change the constraints on the value of $f$ from CMB measurements. This modification can also be understood, as in the previous subsection, from the argument of the effective potential \eqref{eq:f(R,T)_effective_potential}. This type of shift in the mass scale due to $f(R,T)$ gravity will appear in many other similar inflation theories. For instance, the quartic Hilltop inflationary model, presented in \cite{Boubekeur:2005}, considers a potential of the form,
\begin{equation}
V(\varphi) = \Lambda^4 \qty[1 - \qty(\frac{\varphi}{\mu_4})^4 +\ldots]
\end{equation}
where the dots indicate higher-order terms that can be neglected during inflation, and $\mu_4$ is the mass scale that characterizes the inflaton vacuum expectation value, i.e., $\mu_4 \sim \expval{\varphi}$. From the previous analysis of Natural inflation, it is straightforward to show that the action of $f(R,T)$ can be summarized as a correction to the mass scale given by $\mu_4 \to \sqrt{1+2\alpha}\mu_4$. \\

The scalar spectral index and the tensor-to-scalar ratio for the quartic Hilltop model are given, according to \cite{Dimopoulos:2020}, by the following expressions
\begin{subequations}
	\label{eq:Hilltop_GR}
	\begin{align}
	n_{\textrm{S}} - 1 &\approx - \frac{3}{N_t} \qty[\frac{Z}{Z-1}]\\
	r&\approx \frac{128M_{\textrm{Pl}}^8 [4N_tP(Z)]^3 }{\mu_4^{8}  [2(1-ZP(Z))]^2 }  = \frac{8}{3}(1-n_{\textrm{S}}) P(Z), \label{eq:plane_hilltop}
	\end{align}
\end{subequations}
where $N_t$ is the total number of e-folds during inflation, such that 
\begin{equation}
N_t = N + \frac{\mu_4^{2}}{4M_{\textrm{Pl}}^2}
\end{equation}
where $N$ is the number of e-folds after the cosmological scales exit the horizon until the end of inflation, and also
\begin{equation}
Z = 16N_t^2\frac{M_{\textrm{Pl}}^4}{\mu_4^4}, \quad P(Z) = 1-\sqrt{1-\frac{1}{Z}}.
\end{equation}

\begin{figure}[h!]
	{\centering     
		\includegraphics[width = 0.49\textwidth]{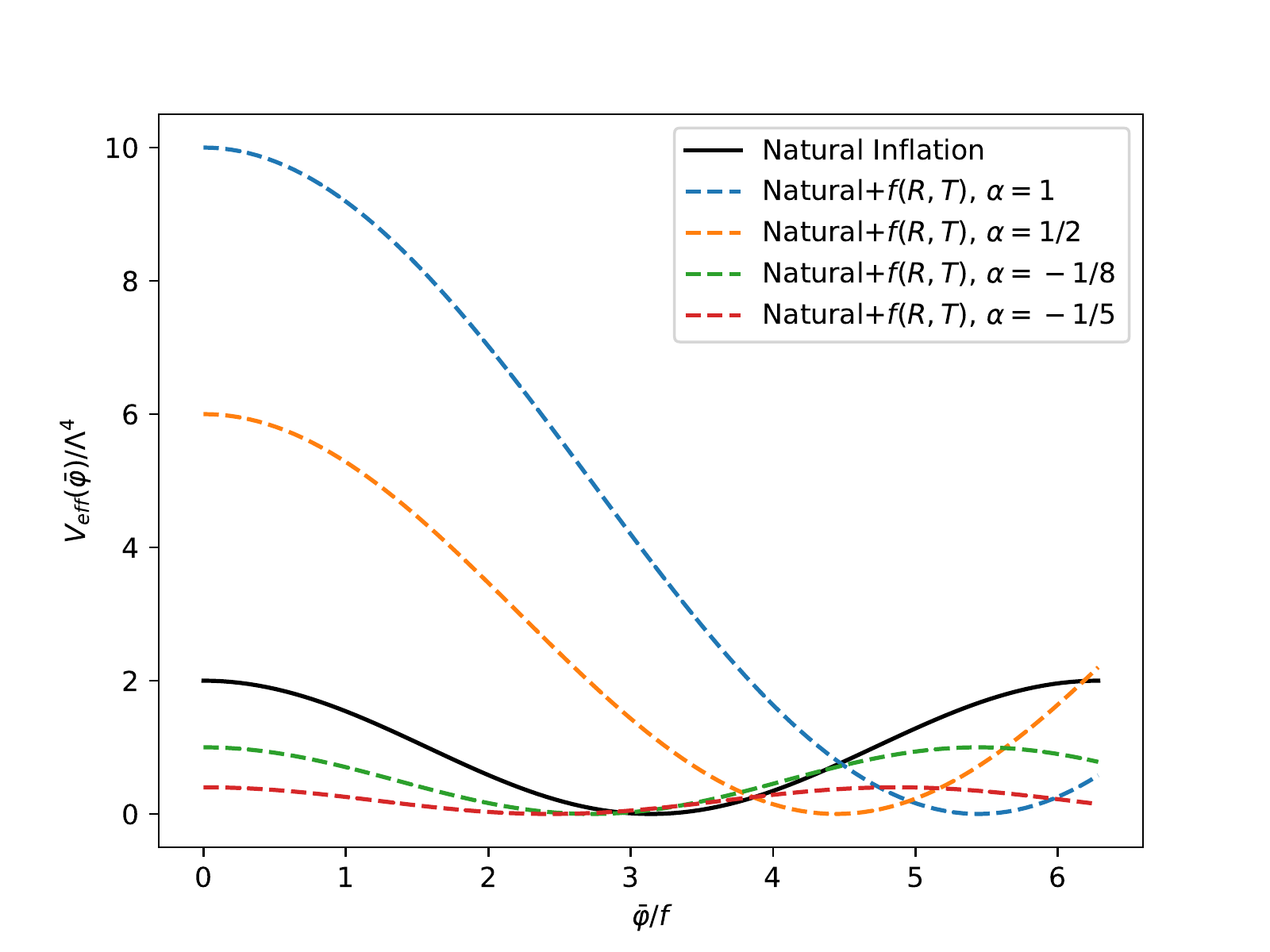}
		\includegraphics[width = 0.49\textwidth]{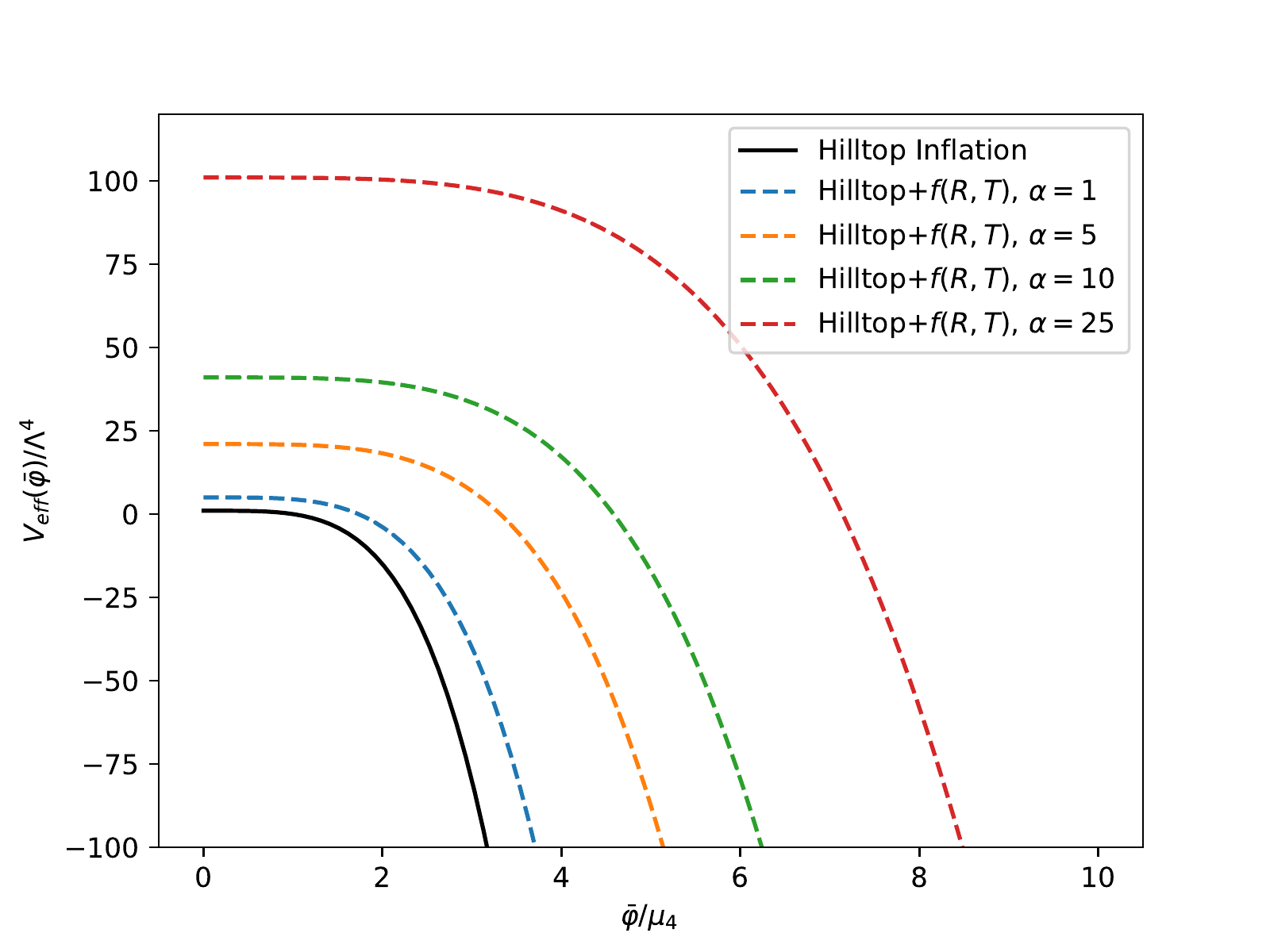} 
	}
	\caption{The solid black lines show the potential of the original Natural/quartic Hilltop inflationary models. In dotted colored lines we have the effective potential, i.e. \eqref{eq:f(R,T)_effective_potential}, for different values of $\alpha$.}
	\label{fig:natural-hilltop}
\end{figure}

By applying the same reasoning as in the previous cases, it is not difficult to prove that when $f(R,T)$ gravity is considered, the induced changes to \eqref{eq:Hilltop_GR} can be expressed as a modification to $Z$, 
\begin{equation}
Z \to \tilde{Z} = \frac{16\tilde{N}_t^2}{(1+2\alpha)^2} \frac{M_{\textrm{Pl}}^4}{\mu_4^4},
\end{equation} 
where $\tilde{N}_t = \tilde{N} + [(1+2\alpha)(\mu_4^2)]/(4M_{\textrm{Pl}}^2)$. Thus, in quartic Hilltop inflation, the action of $f(R,T)$ gravity reduces to a modification of the parameter $\mu_{4}$, as expected from the analysis of the effective potential defined in \eqref{eq:f(R,T)_effective_potential}. In practice, these corrections to $f$ and $\mu_4$ do not modify the shape of the curves in the $(n_{\textrm{S}},r)$ plane, as it can be seen from Fig. \ref{fig:all}. However, the values of $f$ and $\mu_4$ needed to span some region within the $(n_{\textrm{S}},r)$ plane will be different for a non-zero value of $\alpha$. As an example, we show in Table \ref{tab:1} a comparison between the parameters range considered in the analysis of inflationary models done by the Planck collaboration \cite{Planck:2018}, and how they change as we increase the value of $\alpha$. In order to span some fixed region in the $(n_{\textrm{S}},r)$ plane, the accepted range for $\mu_{4}$ changes as we modify the value of $\alpha$. \\
\begin{table}[h!]
	\begin{tabular}{|c|c|c|}
		\hline 
		$\alpha$ & Natural Inflation & Quartic Hilltop Inflation  \\ 
		\hline \hline
		$0$ &  $\phantom{-}0.30<\log_{10}(f/M_{\textrm{Pl}})<2.50$ & $-2.00<\log_{10}(\mu_4/M_{\textrm{Pl}})<2.00$   \\ 
		\hline 
		$1$ &  $\phantom{-}0.06<\log_{10}(f/M_{\textrm{Pl}})<2.26$  & $-2.24<\log_{10}(\mu_4/M_{\textrm{Pl}})<1.76$ \\ 
		\hline 
		$5$ &  $-0.22<\log_{10}(f/M_{\textrm{Pl}})<1.97$ & $-2.52<\log_{10}(\mu_4/M_{\textrm{Pl}})<1.48$ \\ 
		\hline 
		$10$ & $-0.36<\log_{10}(f/M_{\textrm{Pl}})<1.84$ & $-2.66<\log_{10}(\mu_4/M_{\textrm{Pl}})<1.34$\\ 
		\hline 
		$20$ & $-0.51<\log_{10}(f/M_{\textrm{Pl}})<1.69$ & $-2.81<\log_{10}(\mu_4/M_{\textrm{Pl}})<1.19$\\ 
		\hline 
	\end{tabular} 
	\caption{Some examples for the parameter range of Natural/quartic Hilltop inflationary models considering different values of $\alpha$. The $\alpha=0$ cases are the ranges provided in Planck 2018 results \cite{Planck:2018} to successfully span the $(n_{\textrm{S}},r)$ plane within the intervals $r\in [0,0.2]$ and $n_{\textrm{S}}\in[0.93,1.00]$. Natural inflation is strongly disfavored by the data. However, for quartic Hilltop inflation, the corrected constraint of $\mu_4$ to be within the 95\% CL region is given by \eqref{eq:constraint_Hilltop_alfa}.}
	\label{tab:1}
\end{table}

In particular, Natural inflation is strongly disfavored by the Planck 2018 data, and we reach the same conclusion with the contribution of $f(R,T)$ gravity. On the contrary, quartic Hilltop inflation provides a good fit with the data as long as the value of $\mu_4$ is constrained to $\log_{10}(\mu_4/M_{\textrm{Pl}})> 1$ at 95\% CL. Therefore, the contribution of $f(R,T)$ gravity provides the following constraint for $\mu_4$ and $\alpha$ in order to fit the data of Planck 2018,
\begin{equation}
\label{eq:constraint_Hilltop_alfa}
\log_{10}(\mu_4/M_{\textrm{Pl}}) > 1 - \frac{1}{2}\log_{10}(1+2\alpha).
\end{equation}
In general, a higher value of $\alpha$ will decrease the constraint on $\mu_4$. This fact could have some important consequences for the interpretation of $\mu_{4}$. As it can be seen from equation \eqref{eq:constraint_Hilltop_alfa} and as it has been established in previous works, the inflaton \textit{vev} in the quartic Hilltop model should be super--Planckian, i.e. $\mu_{4} > 10M_{\textrm{Pl}}$, to fit the cosmological data. Nevertheless, for values of $\alpha>50$, this is no longer a requirement and actually arises the possibility that $\mu_4 \sim M_{\textrm{Pl}}$ or even $\mu_4\le M_{\textrm{Pl}}$. For instance, if $\tilde{N} \sim 55$, $\mu_{4} \sim M_{\textrm{Pl}}$ and $\alpha \sim 100$, we have $n_{\textrm{S}} \sim 0.9631$ and $r\sim 0.012$, which are in good agreement with the Planck constraints. This kind of corrections could modify the interpretation of the Hilltop model and improve its behavior at the quantum level since it is known that super-Planckian values for the inflaton are problematic from the point of view of particle physics and effective field theory \cite{Mazumdar:2010,Chialva:2014}.\\

In Fig. \ref{fig:natural-hilltop} it can be seen how the shape of effective potential for Natural and Quartic Hilltop Inflation is modified depending on different values of the parameter $\alpha$. In the case of Natural Inflation, the position of the minimum of the potential is shifted to greater values of $\bar{\varphi}/f$ as the value of $\alpha$ increases. Additionally, the height of the plateau in the Hilltop inflationary model is also shifted by changing the value of $\alpha$.

\subsection{Starobinsky Inflation}

The last example we are going to address will be the Starobinsky inflation. The origin lies in Starobinsky's seminal investigations in the early eighties \cite{Starobinsky:1980}. In this model, the standard Einstein--Hilbert action includes a $R^2$ term, 
\begin{equation}
\label{eq:action_starobinsky}
S = \frac{1}{2\kappa} \int \qty(R+ \frac{R^2}{6M^2}) \sqrt{-g} \dd[4]{x},
\end{equation} 
where $M$ is a constant. It is well-known that this model can be recast, by a conformal transformation to the Einstein frame, into a scalar field minimally coupled to gravity \cite{Kaneda:2010,Sebastiani:2015}, i.e., the standard Einstein gravity with a canonically normalized scalar field, $\chi$, and a potential of the form \cite{Piattela:2018}
\begin{equation}
V(\chi) = \frac{3M^2M_{\textrm{Pl}}^2}{4} \qty(1-e^{-\sqrt{2/3} \frac{\chi}{M_{\textrm{Pl}}} })^2.
\end{equation}
Therefore, interpreting the field $\chi$ as our inflaton (also known as a scalaron), we can compute the slow-roll parameters from \eqref{eq:epsilon_V_GR} and \eqref{eq:eta_V_GR},
\begin{align*}
\epsilon_{\textrm{V}} = \frac{4}{3} \frac{e^{2y}}{(1-e^y)^2},\quad \eta_{\textrm{V}} = \frac{4}{3} \frac{(e^y-2e^{2y})}{(1-e^y)^2} ,\quad y = -\sqrt{\frac{2}{3}} \frac{\chi}{M_{\textrm{Pl}}}.
\end{align*}

The number of e--folds can be approximated by calculating the integral \eqref{eq:e-folds_GR}, such that
\begin{equation}
N \approx \frac{3}{4} e^{-y} = \frac{3}{4} e^{\sqrt{\frac{2}{3}} \frac{\chi}{M_{\textrm{Pl}}} }
\end{equation}
Thus, the slow--roll parameters can be written as,
\begin{subequations}
	\begin{align}
	\epsilon_{\textrm{V}} &= \frac{12}{(3-4N)^2} \approx \frac{3}{4N^2} \\
	\eta_{\textrm{V}} &= - \frac{8(2N-3)}{(3-4N)^2} \approx - \frac{1}{N},
	\end{align}
\end{subequations}
wherein each approximation we kept the dominant contributions for large N. Considering the last expressions, we obtain the well-known predictions of the scalar spectral index and the tensor-to-scalar ratio from the Starobinsky model \cite{Piattela:2018},
\begin{subequations}
	\begin{align}
	n_{\textrm{S}} &\approx 1 - \frac{2}{N} \\
	r&\approx \frac{12}{N^2} = 3(1-n_{\textrm{S}})^2
	\end{align}
\end{subequations}
where we used that $\abs{\epsilon_{\textrm{V}}} \ll \abs{\eta_{\textrm{V}}}$ for large $N$. The Starobinsky model has  regained interest in recent years since its predictions seem to be in good agreement with the last CMB measurements from the Planck collaboration \cite{Planck:2018}, but also because it was found a connection between this model and inflation triggered by a Higgs boson non--minimally coupled to gravity \cite{Bezrukov:2007,Kehagias:2013}. As a consequence, many  extensions of the Starobinsky model, like non-local modifications or the inclusion of higher order terms, are currently investigated \cite{Aldabergenov:2018,Cuzinatto:2018,King:2019,Cheong:2020,Koshelev:2020}. \\

A simple question immediately arises: Is it possible to apply $f(R,T)$ gravity to Starobinsky inflation? At first glance, in the Jordan frame, i.e., the action \eqref{eq:action_starobinsky}, 
the model does not consider any matter field, so the trace $T$ is identically zero. Nevertheless, once we are in the Einstein frame, we could apply the prescription of $f(R,T)$ gravity and add a term proportional to the trace of the energy-momentum tensor formed by the inflaton/scalaron. With this procedure, we can follow the previous analysis and compute the corrections to the cosmological parameters due to the additional contribution coming from $\alpha$. \\

\begin{figure}[h!]
	{\centering     
		\includegraphics[width = 0.7\textwidth]{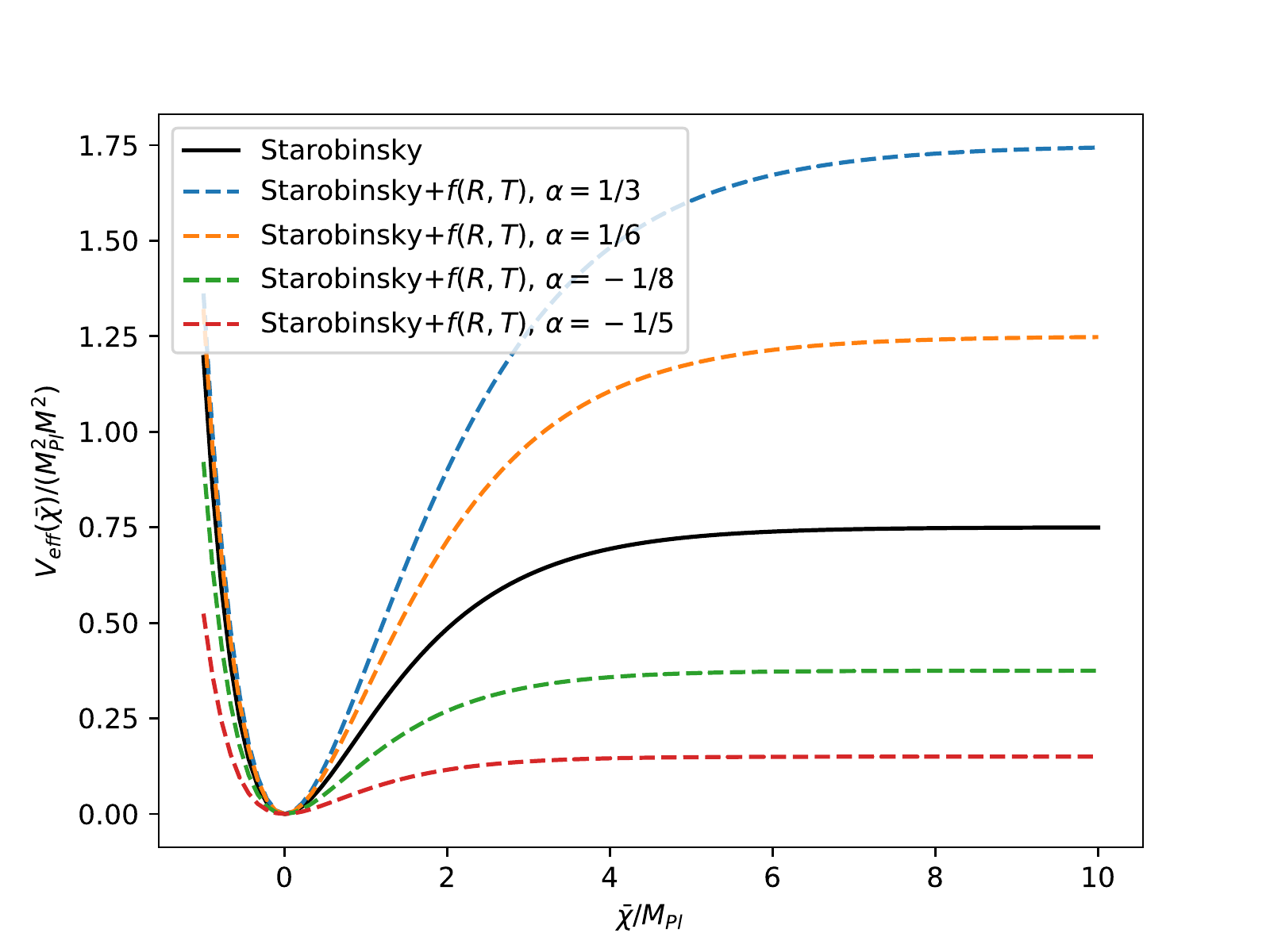}
	}
	\caption{The solid black line shows the original potential in the Einstein frame for Starobinsky inflation. The dotted colored lines illustrate the effective potential for different values of $\alpha$.}
	\label{fig:starobinsky-potential}
\end{figure}

To start, we can analyze the corresponding effective potential for the Starobinsky model, which is illustrated in Fig. \ref{fig:starobinsky-potential}. From the plot, we can observe that the value of $\alpha$ modifies the height of the potential plateau, which can be interpreted as the vacuum energy $V_0$ that dominates the inflation dynamics (for $\chi\gg 1$),
\begin{equation}
V_0 = \frac{3}{4}M_{\textrm{Pl}}^2 M^2 (1+4\alpha).
\end{equation}

The slow-roll parameters can be obtained following the standard procedure, and they read
\begin{subequations}
	\begin{align}
	\tilde{\epsilon}_{\textrm{V}} &= \frac{4}{3} \frac{e^{2y}}{(1+2\alpha)(1-e^y)^2} = \frac{12(1+2\alpha)}{[3(1+2\alpha)-4\tilde{N}]^2} \approx \frac{3(1+2\alpha)}{4\tilde{N}^2}\\
	\tilde{\eta}_{\textrm{V}}&= \frac{4}{3} \frac{(e^y-2e^{2y})}{(1+2\alpha)(1-e^y)^2} = - \frac{8[2\tilde{N} - 3(1+2\alpha) ]}{[3(1+2\alpha) -4\tilde{N}]^2} \approx - \frac{1}{\tilde{N}},
	\end{align}
\end{subequations}
where $\tilde{N}$ is the number of e--folds till the end of inflation, and we kept the contributions from large $N$ as long as $\alpha \ll N$. Therefore, using these slow-roll parameters, we can compute the corrected spectral indices and the tensor-to-scalar ratio,
\begin{subequations}
	\begin{align}
	n_{\textrm{S}} &\approx 1 - \frac{2}{\tilde{N}}\\
	n_{\textrm{T}} &\approx - \frac{3(1+2\alpha)}{2\tilde{N}^2}\\
	r&\approx \frac{12(1+2\alpha)}{\tilde{N}^2} = 3(1+2\alpha)(1-n_{\textrm{S}})^2.
	\end{align}
\end{subequations} 

\begin{figure}[h!]
	{\centering     
		\includegraphics[width = \textwidth]{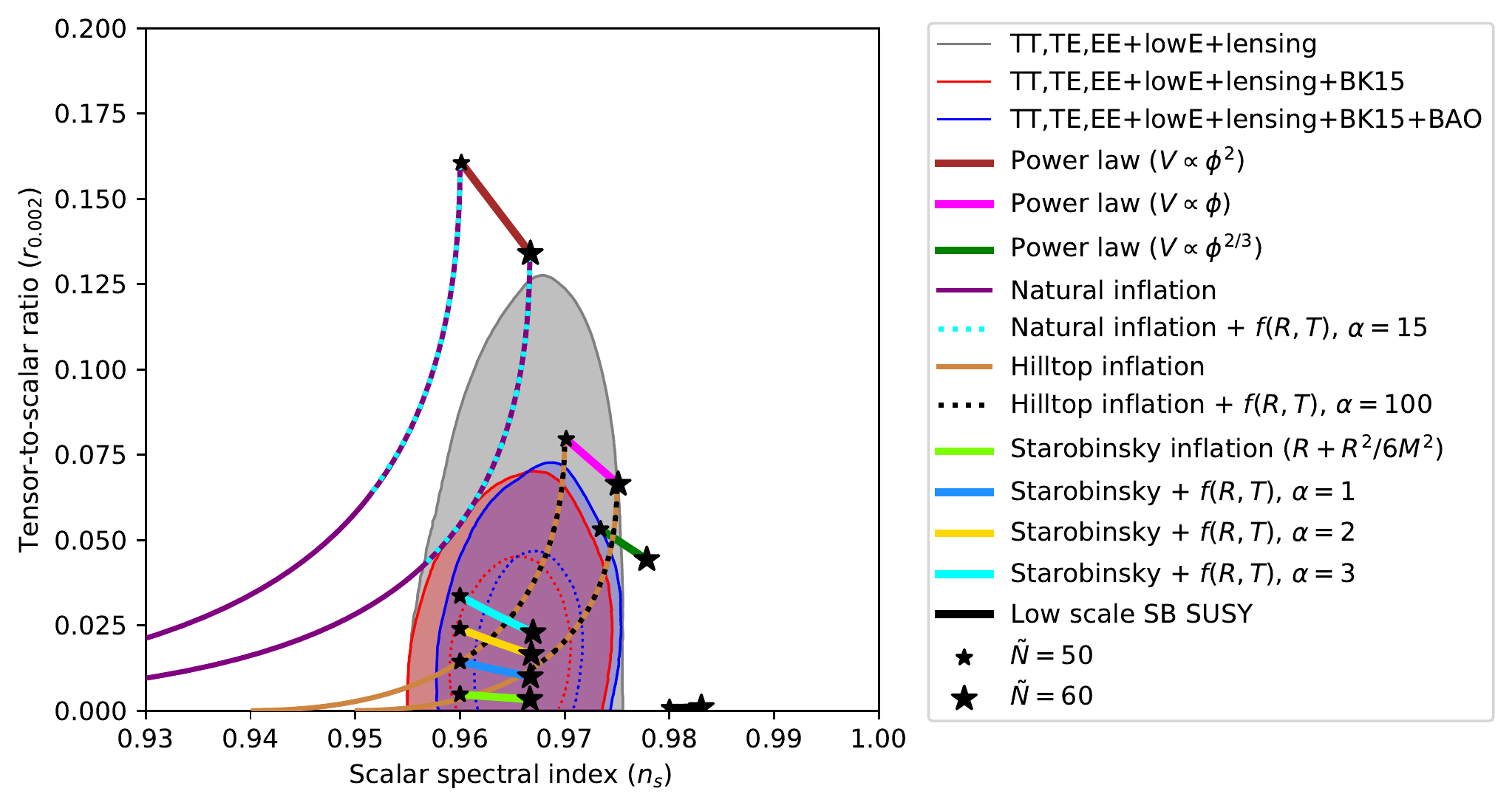}
	}
	\caption{Marginalized joint 68\% (dotted) and 95\% (solid) CL regions for $n_{\textrm{S}}$ and $r$ at $k=0.002$ Mpc$^{-1}$ from Planck 2018 data release \cite{Planck:2018}. We shown the prediction of some selected models and in some cases also the corrections due to $f(R,T)$ gravity are included. We consider that $\tilde{N}$ is the number of e--folds until the end of inflation, according to the modified model, i.e. $\alpha\neq 0$.}
	\label{fig:all}
\end{figure}

For example, $\tilde{N}=55$ and $\alpha = 1$ we have $n_{\textrm{S}} = 0.9636$, $n_{\textrm{T}} = -0.0015$ and $r = 0.012$ (if $\alpha = 0$, $n_{\textrm{T}} = -0.0004$ and $r = 0.004$). To understand how $\alpha$ modifies the trajectories, we plot some examples on the $(n_{\textrm{S}},r)$ plane, including other models, as is illustrated in Fig. \ref{fig:all}. 

Thus, in general, the contribution from $f(R,T)$ gravity will only modify the tensor spectral index's value and the tensor-to-scalar ratio when comparing it to the standard Starobinsky inflation.  For positive values of $\alpha$, the amount of primordial gravitational waves produced during inflation will increase. Hence, if future measurements of the B-modes of the CMB constrain $r$ or $n_{\textrm{T}}$ to values quite different than the predictions of the Starobinsky model, these modifications due to $f(R,T)$ gravity could stand as a viable model. Conversely, for negative values of $\alpha$, the magnitude of $r$ will decrease with respect to the standard prediction. Considering the previous example, for $n_{\textrm{S}} = 0.9636$ and $\alpha = -1/3$, we have $r=0.001$ (compared with $r = 0.004$ if $\alpha = 0$). It should be noted, however, that the negative values of $\alpha$ are bounded from below in order to ensure a well behavior of the physical parameters, i.e. $-1/2 < \alpha$. Moreover, according to the Planck data, the values of the scalar spectral index and the bound on the value of the tensor-to-scalar ratio are given by \cite{Planck:2018},
\begin{subequations}
	\begin{align}
	n_{\textrm{S}} &= 0.9649 \pm 0.0042 \qquad (\textrm{at 68\% CL})\\
	r_{0.002} &< 0.056  \qquad \qquad \qquad \;\: (\textrm{at 95\% CL}).
	\end{align}
\end{subequations}
With this information, we can set a bound on the allowed values of $\alpha$ such that the resulting $r$ and $n_{\textrm{S}}$ are in good agreement with the measurement of Planck, and it is given by
\begin{equation}
-0.5<\alpha < 5.54.
\end{equation}

These are some of the simplest and most popular models, but they are not the only ones. For a comprehensive and extensive compendium of different inflationary models, see~\cite{Martin:2013}.

\section{A brief analysis on models with higher order powers of T}
\label{sec:Non-Minimal_Coupling}

In this manuscript we have considered, the simplest minimal coupling between matter and gravity, i.e.,~$f(R,T) = R+2\kappa \alpha T$. However, it is worth wondering how difficult it is to address a more complex interaction. For instance, let us consider a function of the form $f_{*}(R,T)=R+2\kappa (\alpha T + \beta T^2)$. Using \eqref{eq:f(R,T)_Action_Final}, the equations of motion for this model can be expressed as in \eqref{eq:Effective_T_munu}, with an effective energy--momentum tensor given by,
\begin{equation}
T_{\mu\nu}^{(\textrm{eff*})} = T_{\mu\nu} - 2 (\alpha+2\beta T)(T_{\mu\nu}+\Theta_{\mu\nu})+(\alpha T + \beta T^2)g_{\mu\nu}.
\end{equation}

This is, again, a diagonal tensor, but now with the following components,
\begin{align}
T_{00}^{(\textrm{eff*})} &= \frac{\dot{\varphi}^2}{2} (1+2\alpha + 6\beta \dot{\varphi}^2 -16\beta V) + V(1+4\alpha-16\beta V)\nonumber\\
&\equiv \rho_{\varphi}^{(\textrm{eff*})}, \label{eq:f*T00}\\ T_{ij}^{(\textrm{eff*})} &= \qty( \frac{\dot{\varphi}^2}{2} (1+2\alpha + 2\beta \dot{\varphi}^2 - 16\beta V) - V(1+4\alpha-16\beta V)        )g_{ij}\nonumber\\
& \equiv p_{\varphi}^{(\textrm{eff*})}.  \label{eq:f*Tij}
\end{align}

We can notice that in this example a cross term appears, e.g.,~$\dot{\varphi}^2 V$, in addition to higher order terms like $\dot{\varphi}^4$ or $V^2$. Following a similar procedure as we did with the simplest case, we can write the Friedmann equations in their standard fashion, e.g.,~\eqref{eq:f(R,T)_Friedmann_1}, \eqref{eq:f(R,T)_Friedmann_1}, \eqref{eq:f(R,T)_H_dot}, by doing $\rho_{\varphi}\to\rho_{\varphi}^{(\textrm{eff*})}$ and $p_{\varphi}\to p_{\varphi}^{(\textrm{eff*})}$. Hence, we can define the first slow-roll parameter,
\begin{align}
\tilde{\epsilon}_{*} &= - \frac{\dot{H}}{H^2}\nonumber\\
&= \frac{3\dot{\varphi}^2[1+2\alpha+4\beta(\dot{\varphi}^2-4V)]}{2V(1+4\alpha-16\beta V)+(1+2\alpha-16\beta V)\dot{\varphi}^2+6\beta \dot{\varphi}^4}.
\end{align}

If we impose the slow-roll condition as $\tilde{\epsilon}_{*}\ll 1$, it would reduce to something of the form,
\begin{equation}
\dot{\varphi}^2 (1+2\alpha-16\beta V+3\beta \dot{\varphi}^2) \ll V(1+4\alpha-16\beta V).
\end{equation}

Furthermore, by following the same treatment as before, we can obtain the modified Klein-Gordon equation, which reads
\begin{align}
&\ddot{\varphi} [1+2\alpha-16\beta V +12 \beta\dot{\varphi}^2] + V_{,\varphi} [1+4\alpha-32\beta V+8\beta \dot{\varphi}^2] \nonumber\\
& +3H\dot{\varphi}[1+2\alpha-16\beta V+4\beta \dot{\varphi}^2] = 0.
\end{align}

As we can observe, all the contributions proportional to $\beta$ have crossed or higher order terms, and more important, the coefficients of these crossed terms are different at each stage, e.g.,~the terms $3\beta \dot{\varphi}^4$ and $\beta \dot{\varphi}^4$ in \eqref{eq:f*T00},\eqref{eq:f*Tij}, respectively. Another difficulty can be noticed when we try to define the second slow-roll parameter, because when we derive $\tilde{\epsilon}_{*}$ we obtain,
\begin{equation}
\dot{\tilde{\epsilon_{*}}} = 2H\tilde{\epsilon}_{*}\qty(\tilde{\epsilon}_{*} + \frac{\ddot{\varphi}}{H\dot{\varphi}})
-
\frac{4\dot{\varphi}^2\beta(2V_{,\varphi}-\dot{\varphi}\ddot{\varphi})}{H^2}.
\end{equation}
Then, the second slow-roll parameter is not of the form $\eta \sim \frac{\ddot{\varphi}}{H\dot{\varphi}}$, and the condition $\eta\ll 1$ cannot be used to simplify the expression of $\tilde{\epsilon}_{*}$ in order to construct a simpler slow-roll parameter that depends only on the potential and its derivative, i.e.,~$\tilde{\epsilon}_{V*}$. As a consequence, the slow-roll analysis becomes much more difficult to do in these kind of models, as we cannot easily neglect terms to define more tractable parameters.\\

In summary, since the inclusion of higher order terms of $T$ introduces major corrections on the definition of cosmological parameters, and therefore on the predictions of inflationary models, the study of this type of theories is quite well motivated. However, as it is beyond the scope of this work, we expect to address these issues in future research.

\section{Summary and Conclusions}
\label{sec:discussion}

In this manuscript, we have reviewed the fundamentals of slow-roll inflation in general relativity and investigated the corrections to observable parameters within the context of $f(R,T)$ gravity. We have assumed a model of the form $f(R,T) = R + 2\kappa \alpha T$, where $T$ is the trace of the energy-momentum tensor and $\alpha$ is a constant. This choice has been thoroughly studied in the literature, and usually has been presented as an alternative approach to different cosmological problems, e.g., Dark Energy and Dark Matter. Cosmic inflation, triggered by a perfect fluid or by a simple quadratic potential, was discussed in a recent work \cite{Bhattacharjee:2020_inf}. However, we have focused on a more comprehensive study of the slow-roll approximation, showing some useful and complementary results, which were used in a broad distinct type of inflationary models.\\

In order to study the dynamics of inflation in the $f(R,T)$ theory, we computed the slow-roll parameters, the number of e--folds, the scalar spectral index, the tensor spectral index, and the tensor-to-scalar ratio in the case of a scalar field, $\varphi$, minimally coupled to gravity, with an unknown potential $V(\varphi)$. Thus, we have applied these general results on various models,  e.g., general power-law potentials, Natural Inflation, Quartic Hilltop inflation, and the Starobinsky model. Furthermore, to have a physical intuition of the corrections given by the contributions of $\alpha$, we also shown that the $f(R,T)$ theory is equivalent to a scalar-tensor model of gravity so that we can recast the action to the Einstein frame, i.e., a minimally coupled scalar field in combination with an effective potential. \\

We were able to show the physically significant results by illustrating the trajectories of the models and their corrections on the $(n_{\textrm{S}},r)$ plane, see Fig. \ref{fig:all}. We found that the power-law potentials' trajectories are non affected at all by this specific model of $f(R,T)$ gravity. In the case of Natural and quartic Hilltop inflationary models, the contribution from $f(R,T)$ gravity can modify the constraints on the only parameter of both models, namely $f$ and $\mu_4$. Finally, we applied the prescription of the $f(R,T)$ theory to Starobinsky inflation, where we found that the trajectories in the $(n_{\textrm{S}},r)$ plane can be strongly modified.However, as long as $\alpha$ lies within the interval $-0.5 < \alpha < 5.54$, the trajectories will remain in the 95\% CL region provided by Planck. Moreover, we found that the main contribution of $\alpha$ goes to change the value of $r$, so if the Starobinsky model cannot fit future measurements of primordial gravitational waves, the contribution from the $f(R,T)$ theory could improve the corresponding predictions of the tensor-to-scalar ratio and the tensor spectral index. \\

Finally, we outlined a brief analysis on models with higher order powers of $T$, finding that some crossed terms arise in the equations, making the computation of parameters a harder task. Thus, we leave the analysis of these more complex models of gravity for future research. For instance, a non-minimal coupling between matter and gravity, i.e.,~$f(R, T) \propto RT$, or~$f(R,T)\propto e^{RT}$ could show a different behaviors than the model studied in this work. These alternatives, combined with other types of inflationary models, could complement this work and would provide intriguing results. 

\section*{Acknowledgment}

The author is grateful to R. Velásquez, D. Nicky, A. Mendizábal and N. Padilla for useful comments. M.G. was partially supported by FONDECYT, Chile 1150390 and CONICYT-PIA-ACT14177, and acknowledges the current funding provided by the Pontificia Universidad Católica de Chile and by the Fulbright-Chile Comission through the BIO-CONICYT Doctoral Program.



\end{document}